\documentclass[%
preprint,
superscriptaddress,
 amsmath,amssymb,
]{revtex4-1}

\usepackage[ruled]{algorithm}
\usepackage{algpseudocode}
\algnotext{EndFor}
\algnotext{EndIf}

\usepackage{graphicx}
\usepackage{dcolumn}
\usepackage{bm}
\usepackage[dvipsnames]{xcolor}
\usepackage{amsthm}
\usepackage{float}
\usepackage[normalem]{ulem}
\usepackage{hyperref}

\newtheorem{prop}{PROPOSITION}[section]
\newtheorem{defi}{DEFINITION}[section]

\newtheorem{thm}{THEOREM}[section]

\DeclareMathOperator*{\argmin}{arg\,min} 

\usepackage[caption=false]{subfig}
\usepackage{changes}

\usepackage{hyperref}

\begin{document}




\title{Fast and simple quantum state estimation}

\author{Daniel Uzc\'ategui Contreras}
\affiliation{Departamento de F\'{i}sica, Facultad de Ciencias B\'{a}sicas, Universidad de Antofagasta, Casilla 170, Antofagasta, Chile}
\author{Gabriel Senno}
\affiliation{ICFO-Institut de Ciencies Fotoniques, The Barcelona Institute of Science and Technology, Castelldefels (Barcelona), 08860, Spain}
\author{Dardo Goyeneche}
\affiliation{Departamento de F\'{i}sica, Facultad de Ciencias B\'{a}sicas, Universidad de Antofagasta, Casilla 170, Antofagasta, Chile}

\date{December 14, 2020}

\begin{abstract}
We present an iterative method to solve the multipartite quantum state estimation problem. We demonstrate convergence for any informationally complete set of generalized quantum measurements in every finite dimension. Our method exhibits fast convergence in high dimension and strong robustness under the presence of realistic errors both in state preparation and measurement stages. In particular, for mutually unbiased bases and tensor product of generalized Pauli observables it converges in a single iteration. We show outperformance of our algorithm with respect to the state-of-the-art of maximum likelihood estimation methods both in runtime and fidelity of the reconstructed states.

\end{abstract}

\maketitle
\section{Introduction}
\emph{Quantum state estimation} 
is the process of reconstructing the density matrix from measurements performed over an ensemble of identically prepared quantum systems. In the early days of quantum theory, W. Pauli posed the question of whether position and momentum probability distributions univocally determine the state of a quantum particle \cite{P1933}, something that holds in classical mechanics. However, quantum states belong to an abstract Hilbert space whose dimension exponentially increases with the number of particles of the system. Thus, more information than classically expected is required to determine the state. Since then, it has been having an increasing interest to estimate the state of a quantum system from a given set of measurements and several solutions appeared. For instance, standard state tomography \cite{AJK05} reconstructs $d$-dimensional density matrices from $O(d^3)$ rank-one \emph{Projective Valued Measures (PVM)}, whereas  \emph{mutually unbiased bases (MUB)} \cite{I1981, WF1989} and \emph{Symmetric Informationally Complete (SIC) Positive Operator Valued Measures (POVM)} \cite{RBSC2004} do the same task with $O(d^2)$ rank-one measurement projectors. In general, any tight quantum measurement \cite{SCOTT2006}, equivalently any complex projective $2$-design is informationally complete \cite{H82}.

Quantum state tomography finds applications in communication systems \cite{MVRHZ04}, dissociating molecules \cite{SSJM2003} and characterization of  optical devices \cite{DLPPS2002}. It is a standard tool for verification of quantum devices, e.g. estimating fidelity of two photon CNOT gates \cite{OPWRB2003}, and has been used to characterize quantum states of trapped ions \cite{HHRBCCKR2005}, cavity fields \cite{SDGBRH2012}, atomic ensembles \cite{CBSBOMA2013} and photons \cite{DPS03}. \medskip

Aside from the experimental procedure of conducting a set of informationally complete measurements on a system, quantum tomography requires an algorithm for reconstructing the state from the measurement statistics. From a variety of techniques proposed, the approaches featuring in the majority of experiments are variants of linear inversion (LI) and maximum-likelihood quantum state estimation (MLE) \cite{paris2004quantum}. As its name suggests, with LI one determines the state of the quantum system under consideration by inverting the measurement map solving a set of linear equations with the measurement data as input. For relevant families of informationally-complete set of measurements, analytical expressions for the inverse maps are known, significantly speeding up the whole reconstruction effort, see e.g. \cite{guctua2020fast}. MLE aims to find the state that maximizes the probability of obtaining the given experimental data set, among the entire set of density matrices. Within the different implementations of this basic last idea, those currently achieving the best runtimes are variants of a projected-gradient-descent scheme, see \cite{shang2017superfast,bolduc2017projected}. Algorithms based on variants of linear inversion \cite{kaznady2009numerical,acharya2019comparative} are typically faster than those implementing MLE when the inversion process is taken from already existing expressions \cite{GKKT20}. On the other hand, when restrictions on the rank of the state being reconstructed apply, techniques based on the probabilistic method of compressed-sensing have proven to be very satisfactory \cite{gross2010quantum,cramer2010efficient,acharya2017statistical}. In particular, the statistics based on five rank-one projective measurements is good enough to have high fidelity reconstruct of rank-one quantum states, even under the presence of errors in both state preparation and measurement stages \cite{G15}. It is natural to wonder whether one can find a method achieving fidelities as good as those based on MLE, with markedly better runtimes and without rank restrictions. In this work, we present a general method for quantum state estimation achieving better runtimes and fidelities than the state-of-the-art implementations of MLE.


This paper is organized as follows. In Section \ref{sec:pio}, we introduce the main ingredient of our algorithm: the \emph{Physical Imposition Operator}, a linear operator having an intuitive geometrical interpretation. In Section \ref{sec:algorithm}, we present our iterative algorithm for quantum state estimation based on the physical imposition operator and prove its convergence. In Section  \ref{sec:ultra-convergence} we show that for a wide class of quantum measurements, which include mutually unbiased bases and tensor product of generalized Pauli observables for $N$ qudit systems, convergence is achieved in a single iteration. In Section \ref{sec:simulations}, we numerically study the performance of our algorithm in terms of runtime and fidelity estimation, finding an improvement with respect to the most efficient MLE-based method, as far as we know. Finally, in Section \ref{sec:conclusions} we provide conclusions and future lines of research. Proofs of all our results are presented in Appendix \ref{proofs}.

\section{Imposing physical information}\label{sec:pio}
Consider an experimental procedure $\mathcal{P}$ that prepares a quantum system in some \emph{unknown} state.
Let us  assume that, given some prior knowledge about $\mathcal{P}$, our best guess for $\rho$ is the state $\rho_0$, which could be even the maximally mixed state in absence of prior information. Next, we perform a POVM measurement $A$ composed by $m_A$ outcomes, i.e $A=\{E_i\}_{i\leq m_A}$ on an ensemble of systems independently prepared according to $\mathcal{P}$, obtaining the outcome statistics $\vec{p}=\{p_i\}_{i\leq m_A}$. Given this newly acquired information, 
\begin{quote}
\emph{how can we update $\rho_0$ to reflect our new state of knowledge about the system?} 
\end{quote}
To tackle this question, 
we introduce the \emph{physical imposition operator}, a linear map that replaces the initial predictions about observable $A$ contained in $\rho_0$ with an experimentally observed probability $p_i$. 


\begin{defi}[Physical imposition operator]\label{def:PIOO}
Let $A=\{E_i\}_{i\leq {m_A}}$ be a POVM acting on a $d$-dimensional Hilbert space $\mathcal{H}_d$ and let $\vec{p}\in\mathbb{R}^{m_A}$ be a probability vector. The physical imposition operator associated to $E_i$ and $p_i$ is the linear map 
\begin{equation}\label{def:pio}
T^{p_i}_{E_i}(\rho)=\rho+\frac{(p_i-\mathrm{Tr}[\rho E_i])E_i}{\mathrm{Tr}(E_i^2)},
\end{equation}
for every $i\leq m_A$.
\end{defi}
In order to clarify the meaning of the physical imposition operator (\ref{def:pio}) let us assume for the moment that $A$ is a projective measurement. In such a case, operator $T^{p_i}_{E_i}(\rho)$ takes a quantum state $\rho$, removes the projection along the direction $E_i$, i.e. it removes the physical information about $E_i$ stored in the state $\rho$, and imposes a new projection along this direction weighted by the probability $p_i$. Here, $p_i$ can be either taken from experimental data or simulated by Born's rule with respect to a target state to reconstruct. Note that operator $\rho'=T^{p_i}_{E_i}(\rho)$ reflects the experimental knowledge about the quantum system. As we will show in Section \ref{sec:algorithm}, a successive iteration of PIO along an informationally complete set of quantum measurements allows us to reconstruct the quantum state. For POVM in general, operator (\ref{def:pio}) does not entirely impose the information about the outcome. However, after several imposition of all involved operators PIO the sequence of quantum states successfully converges to a quantum states containing all the physical information, as we demonstrate in Theorem \ref{thm:convergence}.
To simplify notation, along the work we drop the superscript $p_i$ in $T_{E_i}^{p_i}$ when the considered probability $p_i$ is clear from the context.

Let us now state some important facts about PIOs that easily arise from Definition \ref{def:PIOO}. From now on, $\mathfrak{D}(\rho,\sigma):=\mathrm{Tr}[(\rho-\sigma)^2]$ denotes the Hilbert-Schmidt distance between states $\rho$ and $\sigma$.
\begin{prop}\label{prop:pioproperties}
The following properties hold for any POVM $\{E_i\}_{i\leq m_A}$ and any $\rho,\sigma$ acting on $\mathcal{H}_d$:
\begin{enumerate}
\item Imposition of physical information: $\mathrm{Tr}[T^{p_i}_{E_i}(\rho)E_i]=p_i.$
\item Composition: $T^{p_j}_{E_j}\circ T^{p_i}_{E_i}(\rho)=T^{p_i}_{E_i}(\rho)+T^{p_j}_{E_j}(\rho)-\rho-\bigl(p_i-\mathrm{Tr}(\rho E_i)\bigr)\mathrm{Tr}(E_iE_j)E_j/\mathrm{Tr}(E_j)^2.$
\item Non-expansiveness: $\mathfrak{D}(T^{p_j}_{E_j}(\rho),T^{p_j}_{E_j}(\sigma))\leq\mathfrak{D}(\rho,\sigma).$
\end{enumerate}
\end{prop}
Some important observations arise from Prop. \ref{prop:pioproperties}. First, for $j=i$ in the above item \emph{2} we find that 
\begin{equation}\label{projection}
T^2_{E_i}(\rho)=T_{E_i}(\rho),
\end{equation}
for any $\rho$, so operator $T_{E_i}$ is an \emph{orthogonal projection}, for every $i\leq m_A$ and any POVM $\{E_i\}_{i\leq m_A}$. Note that any quantum state $\sigma=T_{E_i}(\rho)$ is a fixed point of $T_{E_i}$, i.e. $T_{E_i}(\sigma)=\sigma $, which simply arises from (\ref{projection}). Roughly speaking, quantum states already having the physical information we want to impose are fixed points of the map $T_{E_j}$. This key property allows us to apply dynamical systems theory \cite{S94} to study the tomographic problem. We consider the alternating projection method, firstly studied by Von Neumann \cite{N49} for the case of two alternating projections and generalized by Halperin to any number of projections \cite{H62}. 

In Theorem \ref{thm:convergence}, we will show that composition of all physical imposition operators associated to an informationally complete set of POVM produces a linear map having a unique attrative fixed point, i.e., the solution to the quantum state tomography problem. The uniqueness of the fixed point guarantees a considerable speed up of the method in practice, as any chosen seed monotonically approaches to the solution of the problem. 

To simplify notation, we consider a single physical imposition operation $\mathcal{T}_A$ for an entire POVM A, defined as follows
\begin{equation}\label{pio2}
\mathcal{T}_A=T_{E_{m_A}}\circ\dots\circ T_{E_1}.
\end{equation}
Up to a constant factor proportional to identity, that we omit, operator $\mathcal{T}_A$ reduces to
\begin{equation}\label{piopvm}
\mathcal{T}_A(\rho)=\sum_{i=1}^{m_A}T_{E_i}(\rho),
\end{equation}
for any PVM $A$, what follows from considering (\ref{pio2}) and Prop.\ref{prop:pioproperties}. This additive property holding for PVM measurements plays an important role, as it helps to reduce the runtime of our algorithm. Precisely, this fact allows us to apply \emph{Kaczmarz method} \cite{K1937} instead of Halpering alternating projection method, for any informationally complete set of PVM. Kaczmarz method considers projections over the subspace generated by the intersection of all associated hyperplanes, defined by the linear system of equations (Born's rule).

Let us introduce another relevant concept
\begin{defi}[Generator state]
Given a POVM $A=\{E_i\}_{i\leq m_A}$ and a probability vector $\vec{p}\in\mathbb{R}^{m_A}$, a quantum state $\rho_{gen}$ is called \emph{generator state} for $\vec{p}$ if $\mathrm{Tr}(\rho_{gen} E_i)=p_i$, for every $i\leq m_A$ . 
\end{defi}
Note that $\rho_{gen}$ is a fixed point of $\mathcal{T}_{E_i}$, according to (\ref{pio2}) and Prop. \ref{prop:pioproperties}. State $\rho_{gen}$ plays an important role to implement numerical simulations, as it guarantees to generate sets of probability distributions compatible with the existence of a positive semidefinite solution to the quantum state tomography problem.

\begin{figure}[ht]
      \includegraphics[scale=0.5]{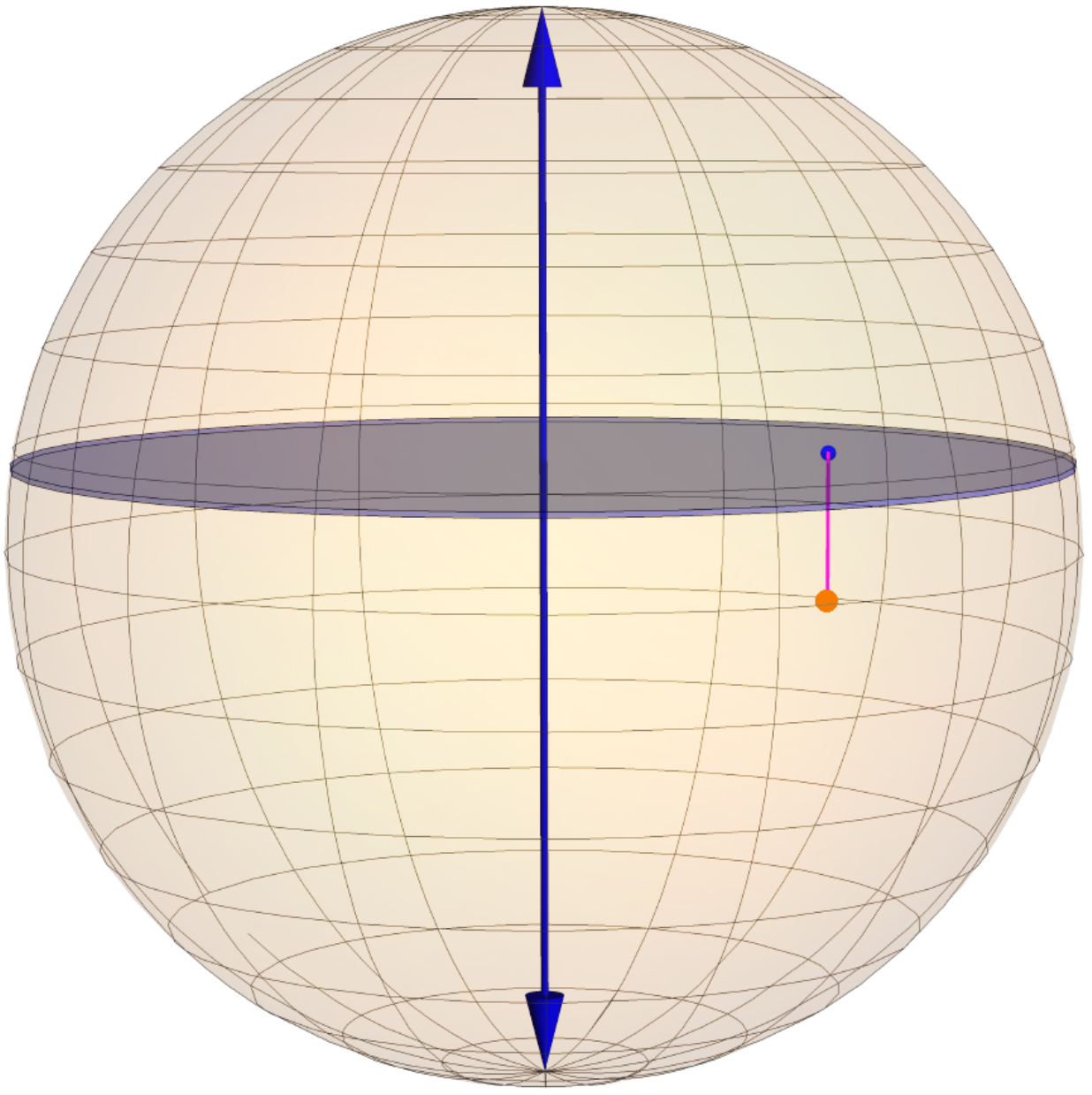}
     \caption{Bloch sphere representation for a single qubit system and PVM measurements. The blue arrows define eigenvectors of $\sigma_z$. The disk shown represents the entire set of quantum states $\rho_{gen}$ satisfying equations $p_j=\mathrm{Tr}(\rho_{gen}E_j)$, $j=0,1$, where $\{E_j\}$ is the set of rank-one eigenprojectors of an observable and $\{p_j\}$ the set of probabilities experimentally obtained. The action of $\mathcal{T}_A$ over the initial state $\rho_0$ (orange dot) is the orthogonal projection to the plane that contains the disk (blue dot) [color online].}
     \label{Fig1}
\end{figure}

To end this section, note that map $\mathcal{T}_A$ defined in (\ref{piopvm}) has a simple interpretation in the Bloch sphere for a qubit system, see Fig. \ref{Fig1}. The image of $\mathcal{T}_A$, i.e. $\mathcal{T}_A[\textrm{Herm}(\mathcal{H}_2)]$, is a plane that contains the disk $$D^{\vec{p}}_{A}=\{z=p_2-p_1\mid z=\mathrm{Tr}(\rho \sigma_z),\,p_i=\mathrm{Tr}(\rho E_i),\,\rho\geq0,\mathrm{Tr}(\rho)=1\},$$ i.e., the disk contains the full set of generator states $\rho_{gen}$. Note that $\mathcal{T}_A$ is not a completely positive trace preserving (CPTP) map, as $\mathcal{T}_A[\textrm{Herm}(\mathcal{H}_2)]$ extends beyond the disk $D^{\vec{p}}_{A}$, i.e. outside the space of states. Indeed, for any state $\rho$ that is not a convex combination of projectors $E_i$, there exists a probability distribution $\vec{p}$ such that $\mathcal{T}_A(\rho)$ is not positive semi-definite. Roughly speaking, any point inside the Bloch sphere from Fig. \ref{Fig1} but outside the blue vertical line is projected by $\mathcal{T}_A$ outside the sphere, for a sufficiently small disk $D^{\vec{p}}_A$.\medskip

\section{Algorithm for quantum state estimation}\label{sec:algorithm}
In the practice of quantum state tomography, one collects a set of probability distributions $\vec{p_1},\dots,\vec{p_{\ell}}$ from a set of $\ell$ POVM measurements $A_1=\{E_i^{(1)}\}_{i\leq m_1},\dots,A_{\ell}=\{E_i^{(\ell)}\}_{i\leq m_{\ell}}$, implemented over an ensemble of physical systems identically prepared in a quantum state $\rho_{gen}$. The statistics collected allows a unique state reconstruction when considering an  \emph{informationally-complete} (IC) sets of observables $A_1,\dots,A_\ell$. Our algorithm for quantum state estimation, Algorithm \ref{alg:pio1} below, defines a sequence of hermitian operators $\rho_n$, not necessarily composed by quantum states, that converges to the unique quantum state that is solution to the tomography problem, i.e. $\rho_{gen}$. For the moment, we assume error-free state tomography in our statements. The algorithm applies to any finite dimensional Hilbert space $\mathcal{H}_d$, and any informationally complete set of quantum observables.
\begin{algorithm}[H]\caption{Quantum state estimation algorithm.}\label{alg:pio1}

\begin{algorithmic}
\vspace{0.2cm}
\Require dimension $d\in\mathbb{N}$, POVMs $A_1,\dots, A_{\ell}$ acting on $\mathcal{H}_{d}$, \\ \hspace{1cm} experimental frequencies $\vec{f}_1,\dots,\vec{f}_{\ell}\in \mathbb{R}^m$ and accuracy $\epsilon\in [0,1]$.
\Ensure estimate $\rho_{\rm est}\in\mathcal{B}(\mathcal{H}_d)$.
\State{$\rho_{0} = \mathbb{I}/d$}
\State{$\rho = \mathcal{T}_{A_{\ell}}\circ\cdots\circ \mathcal{T}_{A_1}(\rho_{0})$}
\Repeat{\\
\hspace*{0.5cm}$\rho_{\rm old} = \rho$\\
\hspace*{0.5cm}$\rho = \mathcal{T}_{A_{\ell}}\circ\cdots\circ \mathcal{T}_{A_1}(\rho_{\rm old})$}
\Until{ $\mathfrak{D}(\rho,\rho_{\rm old})\leq\epsilon$ }\\
\Return{ $\argmin_{ \rho_{\rm est} \in \mathcal{ D(\mathcal{H}_{\rm d} }) } \mathfrak{D}(\rho,\rho_{\rm est})$ } 

\end{algorithmic}
\end{algorithm}

\medskip

In Algorithm \ref{alg:pio1}, $\mathcal{ D( \mathcal{H}_{\rm d} ) }$ denotes the set of density operators over $\mathcal{H}_{\rm d}$. Theorem \ref{thm:convergence} below asserts the convergence of Algorithm \ref{alg:pio1} when the input frequencies are exact, i.e. Born-rule, probabilities of an IC set of POVMs.

\begin{thm}\label{thm:convergence}
Let $A_1,\dots,A_{\ell}$ be a set of  informationally complete POVMs acting on a Hilbert space $\mathcal{H}_d$, associated to a compatible set of probability distributions $\vec{p_1},\dots,\vec{p_{\ell}}$. Therefore, Algorithm \ref{alg:pio1} converges to the unique solution to the quantum state tomography problem.
\end{thm}
Here, compatibility refers to the existence of a quantum state associated to exact probability distributions $\vec{p_1},\dots,\vec{p_{\ell}}$ what is guaranteed when probabilities come from a generator state $\rho_{gen}$. Theorem \ref{thm:convergence} asserts that the composite map $\mathcal{T}_{A_{\ell}}\circ\cdots\circ \mathcal{T}_{A_1}$ defines a dynamical system having a unique attractive fixed point. The successive iterations of Algorithm \ref{alg:pio1} define a \emph{Picard sequence} \cite{KCG90}:
\begin{align}\label{eq:picard-sequence}
    \rho_0&=\mathbb{I}/d,\nonumber\\
    \rho_n&= \mathcal{T}_{A_{\ell}}\circ\dots\circ \mathcal{T}_{A_1}(\rho_{n-1}),~n\geq 1.
\end{align} 
Note that for arbitrary chosen set of observables, the composition of physical imposition operators depends on its ordering. According to Theorem \ref{thm:convergence}, this ordering does not affect the success of the convergence in infinitely many steps. However, in practice one is restricted to a finite sequence, where different orderings produce different quantum states as an output. Nonetheless, such difference tends to zero when the state $\rho_n$ is close to the attractive fixed point, i.e. solution to the state tomography problem. According to our experience from numerical simulations, we did not find any advantage from considering a special ordering for composition of operators.

\begin{figure}[t]
      \includegraphics[scale=0.5]{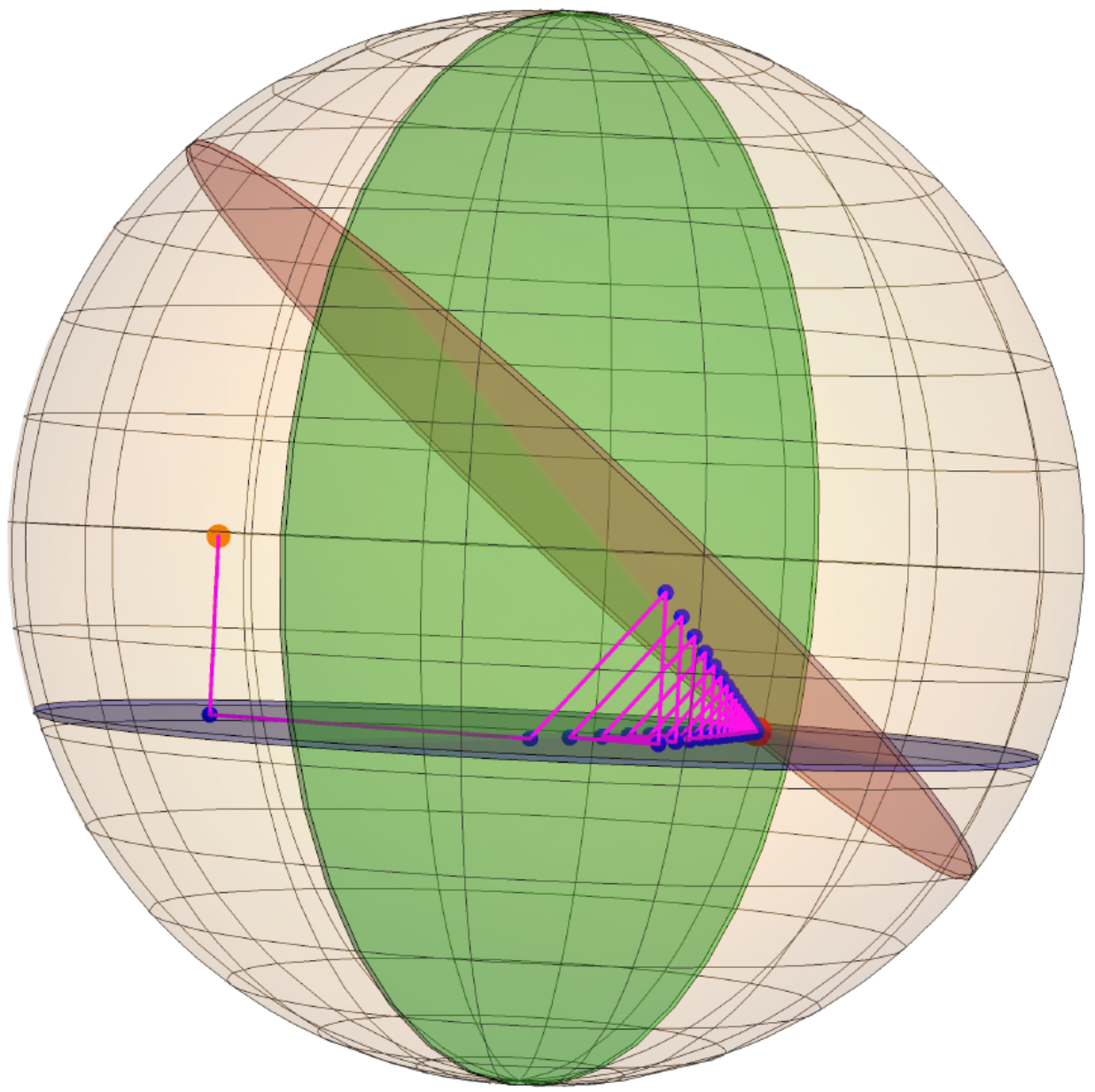}
     \caption{Graphical representation of the convergence of Algorithm \ref{alg:pio1} in the Bloch sphere for a single qubit system. We show convergence for three incompatible PVMs $A_1,A_2$ and $A_3$, defining disks $D_1$ (grey), $D_2$ (green) and $D_3$ (red) on the Bloch sphere. The initial state $\rho_0$  (orange dot), which we have chosen different from $\mathbb{I}/2$ only for graphical purposes, is first projected to $D_1$. The corresponding point in $D_1$ is then projected to $D_2$ and that projection is later projected to $D_3$. The iteration of this sequence of projections successfully converges to the generator state $\rho_{gen}$ (red dot), the unique solution to the quantum state tomography problem [color online].} 
     \label{Fig2}
\end{figure}

 Figure \ref{Fig2} shows the convergence of $\rho_n$ in the Bloch sphere representation for a single qubit system and three PVMs taken at random. For certain families of measurements, e.g. mutually unbiased bases and tensor product of Pauli matrices, the resulting Picard sequences and, therefore, Algorithm \ref{alg:pio1} converge in a single iteration, see Prop. \ref{prop_singlestep}. That is, $\rho_n = \rho_{1}$ for every $n\geq 1$. We numerically observed this same behaviour for the $3^N$ product Pauli eigenbases in the space of $N$-qubits, with $1\leq N\leq 8$, conjecturing that it holds for every $N\in\mathbb{N}$, see Section \ref{sec:simulations-pauli}. 








In a previous work \cite{GdlT2014}, a related algorithm was introduced for quantum state estimation. However, it has several disadvantages with respect to our work, namely: (\emph{i}) it works for pure states only; (\emph{ii}) the dynamics is non-linear, requiring a large runtime to converge (\emph{iii}) convergence to the target state is not guaranteed. The main reason behind this last property is the existence of a large amount of undesired basins of attraction, as the solution to the problem is not the only attractive fixed point; finally, (\emph{iv}) realistic state reconstruction is not possible due to the impossibility to introduce realistic noise, as it destroys purity. Note that Algorithm \ref{alg:pio1} does not reduce to the one defined in Ref. \cite{GdlT2014} when reconstructing pure states, as our imposition operator is linear.

\subsection{Ultra-fast convergence}\label{sec:ultra-convergence}
When considering maximal sets of mutually unbiased bases, the Picard sequences featuring in Algorithm \ref{alg:pio1} converge in a single iteration. This is so because the associated imposition operators commute for MUB. This single-iteration convergence is easy to visualize in the Bloch sphere for a qubit system, as the three disks associated to three MUB are mutually orthogonal, and orthogonal projections acting over orthogonal planes keep the impositions within the intersection of the disks. The same argument also holds in every dimension. Let us formalize this result.
\begin{prop}\label{MUBcommute}
Let $T_A$ and $T_B$ be two physical imposition operators associated to two mutually unbiased bases $A$ and $B$. Therefore,
\begin{equation}
T_B\circ T_A=T_A\circ T_B=T_A+T_B-\mathbb{I}.
\end{equation}
In particular, note that $T_A$ and $T_B$ commute.
\end{prop}


Also, it is easy to see from Item \emph{2}, Prop. \ref{prop:pioproperties} that operators $T_{E_i}$ commute when considering $E_i$ equal to the tensor product local Pauli group. In this case, operators $E_i$ do not form a POVM but given that they define an orthogonal basis in the matrix space, they are an informationaly complete set of observables.  Let us now show the main result of this section:
\begin{prop}\label{prop_singlestep}
Algorithm \ref{alg:pio1} converges in a single iteration to the unique solution of the quantum state tomography problem for  product of generalized Pauli operators and also for $d+1$ mutually unbiased bases, in any prime power dimension $d$.
\end{prop}



We observe from simulations that the speedup predicted by Prop. \ref{prop_singlestep} has no consequences in the reconstruction fidelity of our method, which is actually higher than the one provided by MLE.

\section{Numerical study}\label{sec:simulations}
Theoretical developments from Sections \ref{sec:pio} and \ref{sec:algorithm} apply to the ideal case of error free probabilities coming from an exact generator state $\rho_{gen}$. In practice, probabilities are estimated from frequencies, carrying errors due to finite statistics. Moreover, the states being prepared in each repetition of the experiment are affected by unavoidable systematic errors. These sources of errors imply that the output of Algorithm \ref{alg:pio1} is typically outside the set of quantum states when considering experimental data. We cope with this situation by finding the closest quantum state to the output, called $\rho_{\rm est}$ in Hilbert-Schmidt (a.k.a. Frobenius) distance , for which there are closed-form expressions \cite{guctua2020fast}. 
In the following, we provide numerical evidence for robustness of our method in the finite-statistics regime with white noise affecting the generator states, i.e. errors at the preparation stage. That is, we consider noisy states of the form $\tilde{\rho}(\lambda)=(1-\lambda)\rho+\lambda\mathbb{I}/d$, where $\lambda$ quantifies the amount of errors. We understand there are more sophisticated techniques to consider errors, e.g. ill-conditioned measurement matrices \cite{bolduc2017projected}. Nonetheless, we believe the consideration of another model to simulate a small amount of errors would not substantially change the exhibited results. 
We reconstruct the state for $N$-qubit systems with $1\leq N\leq 8$, by considering the following sets of measurements: a) Mutually unbiased bases, b) Tensor product of local Pauli bases and c) A set of $d+1$ informationally complete bases taken at random with Haar distribution. The last case does not have a physical relevance but illustrates performance of our algorithm for a set of measurements defined in an unbiased way. As a benchmark, we compare the performance of our method with the conjugate gradient, accelerated-gradient-descent (CG-AGP) implementation of Maximum Likelihood Estimation (MLE) \cite{shang2017superfast}. Computations were conducted on an Intel core i5-8265U laptop with 8gb RAM. For the CG-AGP algorithm, we used the implementation provided by authors of Ref. \cite{shang2017superfast}, see Ref. \cite{superfast-implementation}. We provide an implementation of our Algorithm \ref{alg:pio1} in Python \cite{pio-implementation}, together with the code to run the simulations presented in the current section.

\subsection{Mutually unbiased bases}\label{sec:simulations-MUB}
\begin{figure}
     \subfloat[\label{Fig3a}]{%
       \includegraphics[scale=0.51]{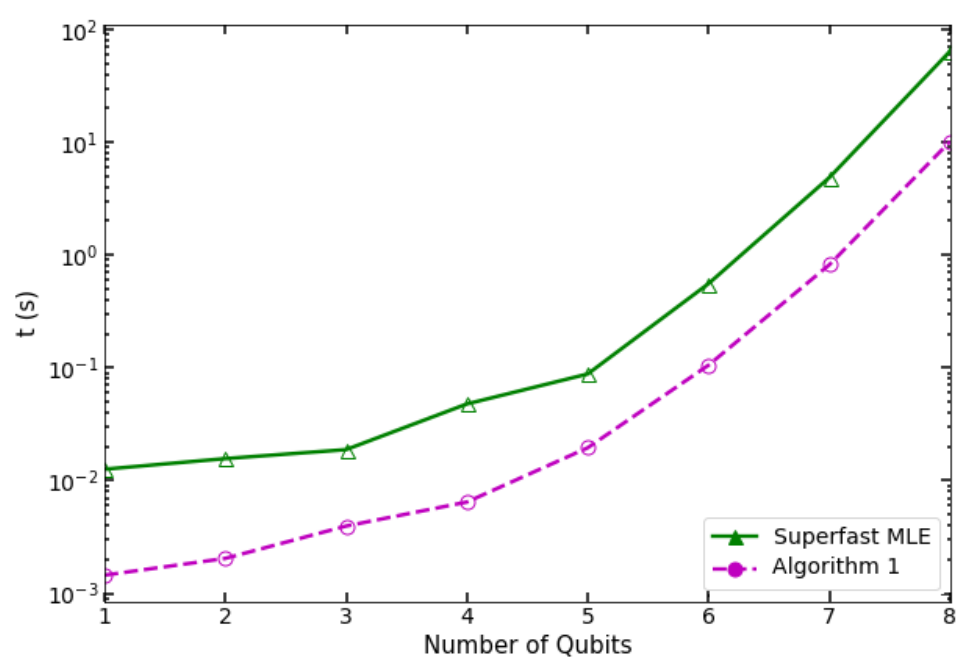}
     }
     \subfloat[\label{Fig3b}]{%
       \includegraphics[scale=0.51]{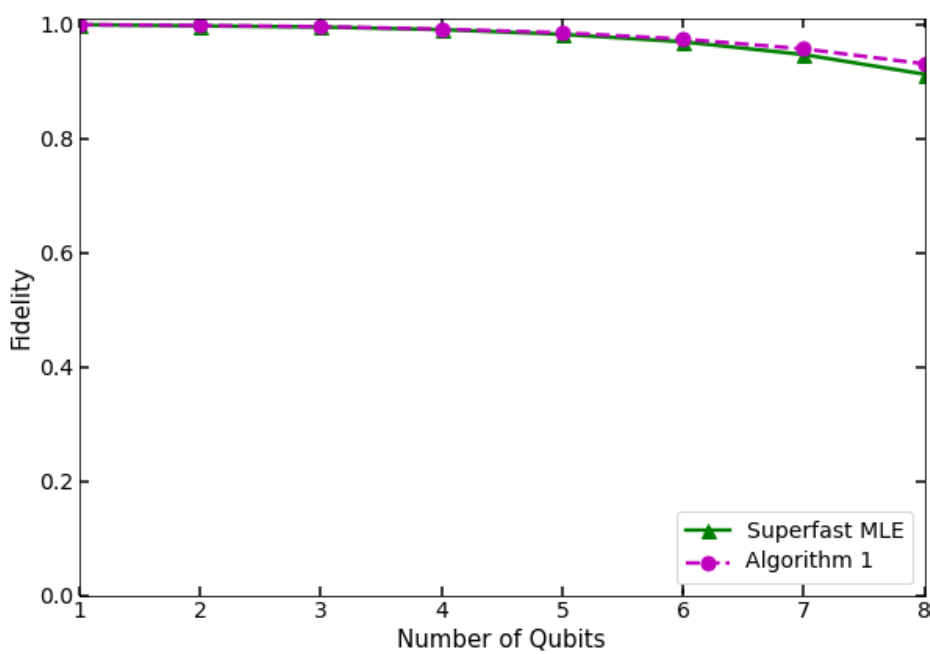}
     }
     \caption{Performance of Algorithm 1
     and the CG-AGP Super-Fast MLE method from \cite{shang2017superfast}, for the reconstruction of $N$-qubit states from a maximal set of $d+1=2^N+1$ mutually unbiased basis (MUB) in dimension $d=2^N$. Generator state $\rho$ is chosen at random by considering the Haar measure distribution, subjected to $10\%$ of white noise and finite statistics satisfying Poissonian distribution. For simulations we consider  $100\times2^N$ samples. Fig.(\ref{Fig3a}) considers runtime of the algorithm in seconds, averaged over 50 trials, whereas (\ref{Fig3b}) shows fidelity between the target and obtained state, also averaged over 50 trials. Despite our runtime is about 1 order of magnitud faster than the Super-Fast MLE, it is worth to mention that we consider simulations in Python and Ref. \cite{shang2017superfast} considers Matlab, so it is not fair to conclude that our algorithm is faster.}     \label{fig:MUB}
\end{figure}

Figure \ref{fig:MUB} shows performance of Algorithm \ref{alg:pio1} in the reconstruction of $N$-qubit density matrices from the statistics of a maximal set of $2^N+1$ MUBs. We consider a generator state $\rho_{gen}$ in dimension $d$, taken at random according to the Haar measure distribution, with the addition of a $10\%$ level of white noise, i.e.
$\tilde{\rho}(\lambda)=(1-\lambda)\rho+\lambda\mathbb{I}/2^N$, with $\lambda=0.1$. Here, it is important to remark that fidelities are compared with respect to the generator state $\rho_{gen}$, so that the additional white noise reflects the presence of systematic errors in the state preparation process. Probabilities are estimated from frequencies, i.e. $f_j=\mathcal{N}_j/\mathcal{N}$ with $\mathcal{N}_j$ the number of counts for outcome $j$ of some POVM and $\mathcal{N}=\sum_j \mathcal{N}_j$ the total number of counts.
 Our simulations consider $\mathcal{N}=100\times 2^N$ samples per measurement basis. 
Our figure of merit is the fidelity $F(\rho_n,\rho_{gen})=\mathrm{Tr}{\sqrt{\sqrt{\rho_{gen}}\rho_n\sqrt{\rho_{gen}}}}^2$ between the reconstructed state after $n$ iterations $\rho_n$ and the generator state $\rho_{gen}$. Runtime of the algorithm is averaged over 50 independent runs, each of them considering a generator state $\rho_{gen}$ chosen at random according to the Haar measure.

\subsection{$N$-qubit Pauli bases}\label{sec:simulations-pauli}
Here, we consider the reconstruction of $N$-qubit density matrices from the $3^N$ PVMs determined by all the products of single qubit Pauli eigenbases, for $N=1,\dots,8$. Similarly to the case of MUBs, Picard sequences $\rho_n=T_{Pauli}^n(\rho_0)$ converge in a single iteration when product of Pauli measurements are considered, for any generator state $\rho_{gen}$ and any initial state $\rho_0$. Figure \ref{fig:pauli} shows performance of a single iteration of these Picard sequences, where the generator state $\rho_{gen}$ is taken at random, according to the Haar measure. Algorithm CG-AGP exploits the product structure of the $N$-qubit Pauli bases to speedup its most computationally expensive part: the computation of the probabilities given by the successive estimates in the MLE optimization. It does so by working with reduced density matrices which, in turn, imply an efficient use of memory. In order to have a fair comparison with our method, 
we decided to include the time to compute the $N$-qubit observables from the single Pauli observables in the total runtime of our algorithm. In practice, however, one would preload them into memory, as they are, of course, not a function of the input, i.e. of the observed probabilities. 
%
%
Nonetheless, Fig. \ref{fig:pauli} shows that our Algorithm \ref{alg:pio1} has a considerable reduction of runtime and better fidelities with respect to the algorithm provided in Ref. \cite{shang2017superfast}.

\subsection{Random measurements for $N$-qubit systems}\label{sec:simulations-random}
\begin{figure}[h!]
     \subfloat[\label{fig:pauli-a}]{%
       \includegraphics[scale=0.38]{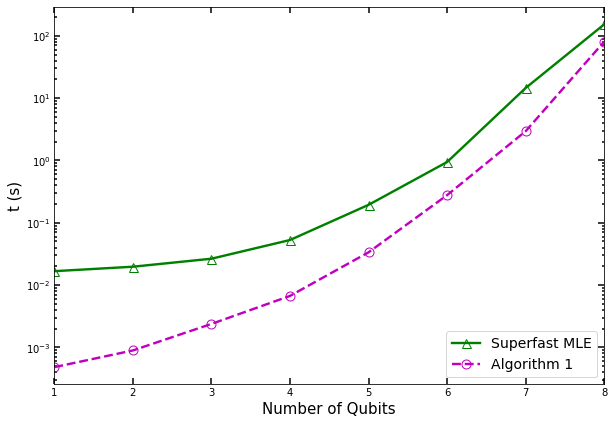}
     }
     \subfloat[\label{fig:pauli-b}]{%
       \includegraphics[scale=0.38]{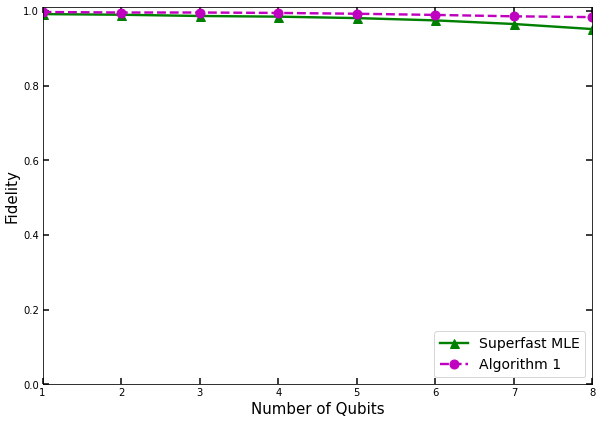}
     }
     \caption{Performance of Algorithm \ref{alg:pio1} and the CG-AGP Super-Fast MLE \cite{shang2017superfast}, for the reconstruction of $N$-qubit states from $3^N$ PVM given by products of the eigenbases of local Pauli observables $\sigma_X$, $\sigma_Y$ and $\sigma_Z$. Generator states $\rho$ are chosen at random (Haar measure), subjected to $10\%$ of white noise and finite statistics satisfying Poissonian distribution, considering  $500\times2^N$ samples per PVM. Fig(\ref{fig:pauli-a}) considers runtime of the algorithm in seconds, whereas (\ref{fig:pauli-b}) shows fidelity between the target stata $\rho$ and reconstructed state, averaged over 50 trials in both cases. We consider simulations in Python, whereas Ref. \cite{shang2017superfast} considers Matlab, so it is not fair to conclude that our algorithm is faster.}
     \label{fig:pauli}
\end{figure}

\begin{figure}[h!]
     \subfloat[\label{fig:random-a}]{%
       \includegraphics[scale=0.38]{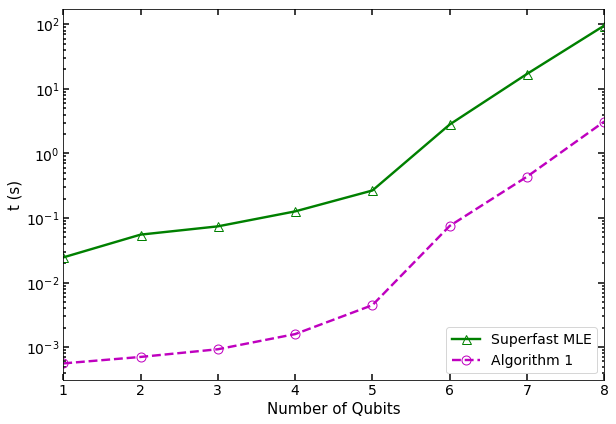}
     }
     \subfloat[\label{fig:random-b}]{%
       \includegraphics[scale=0.38]{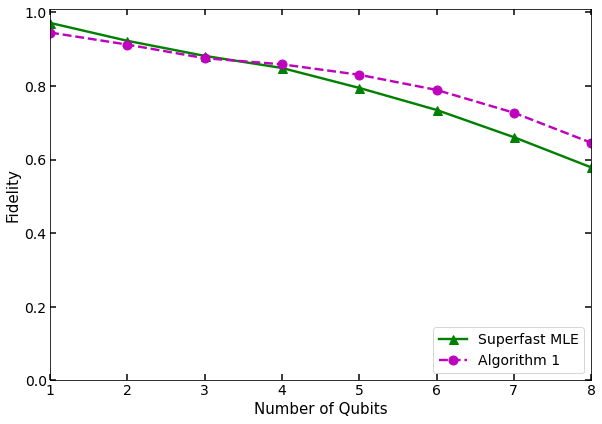}
     }
     \caption{Performance of Algorithm \ref{alg:pio1} and the CG-AGP Super-Fast MLE method from \cite{shang2017superfast} for the reconstruction of $N$-qubit states from a a set of $d+1=2^N+1$ basis chosen Haar-random in dimension $d=2^N$. Algorithm \ref{alg:pio1} was run for $25$ steps or until the Hilbert-Schmidt distance between successive iterates was below $\epsilon=10^{-6}$, whichever happens first.  Generator state $\rho$ is chosen at random by considering the Haar measure distribution, subjected to $10\%$ of white noise. Measurement statistics are estimated from $\mathcal{N}=100\times2
    ^N$ identical copies. Fig(\ref{fig:random-a}) considers the runtime of the algorithm in seconds, averaged over 50 trials, whereas (\ref{fig:random-b}) shows the fidelity between the target and obtained state, also averaged over 50 trials. We consider simulations in Python, whereas Ref. \cite{shang2017superfast} considers Matlab, so it is not fair to conclude that our algorithm is faster.}
     \label{fig:random-bases}
\end{figure}

The simulations in the preceding subsections correspond to informationally complete sets of measurements for which Algorithm \ref{alg:pio1} converges in a single iteration. To test whether the advantage over \cite{shang2017superfast} hinges critically on this fact, we have numerically tested our algorithm with sets of PVMs selected at random, with respect to the Haar measure. In Fig. \ref{fig:random-bases} we show that in this case, the advantage fidelity increases substantially, compared to Figs. \ref{fig:MUB} and \ref{fig:pauli}. 

Finally, we would like to mention the \emph{Projective Least Squares} (PLS) quantum state reconstruction \cite{GKKT20}. This method outperforms both in runtime and fidelity our Algorithm \ref{alg:pio1}. This occurs when the linear inversion procedure required by the method \emph{is not} solved but taken from analytically existing reconstruction formula. Existing inversion formulas are known for to complex projective 2-designs, measurement composed by stabilizer states, Pauli observables and uniform/covariant POVM, see \cite{GKKT20}. However, when taking into account the cost of solving the linear inversion procedure, our method has a considerable advantage over PLS. For instance, PLS does not have such efficient speed up for a number of physically relevant observables for which there is no explicit inversion known, including the following cases: a) discrete Wigner functions reconstruction for arbitrary dimensional boson and fermions quantum systems from discrete quadratures, that be treated as observables by considering Ramsey techniques \cite{L96}, b) reconstruction of single quantized cavity mode from magnetic dipole measurements with Stern-Gerlach aparatus \cite{WCZ96}, c) minimal state reconstruction of $d$-dimensional quantum systems from POVM consisting on $d^2$ elements, inequivalent to SIC-POVM \cite{W06}, d) spin $s$ density matrix state reconstruction from Stern-Gerlach measurements \cite{WA99}, e) Quantum state tomography for multiparticle spin $1/2$ systems \cite{DMP03}, neither reduced to mutually unbiased bases nor local Pauli measurements.

\section{Discussion and conclusions}\label{sec:conclusions}
We introduced an iterative method for quantum state estimation of density matrices from any informationally complete set of quantum measurements in any finite dimensional Hilbert space. We demonstrated convergence to the unique solution for any informationally complete or overcomplete set of POVMs, see Theorem \cite{DMP03}. The method, based on dynamical systems theory, exhibited a simple and intuitive geometrical interpretation in the Bloch sphere for a single qubit system, see Figs. \ref{Fig1} and \ref{Fig2}. Our algorithm revealed an ultra-fast convergence for a wide class of measurements, including mutually unbiased bases and tensor product of generalized Pauli observables for an arbitrary large number of particles having $d$ internal levels. These results considerably improved both the runtime and fidelities reported by the CG-AGP Super-Fast MLE estimation \cite{shang2017superfast} for all the studied cases, see Section \ref{sec:ultra-convergence}. Furthermore, numerical simulations revealed strong robustness under the presence of realistic errors in both state preparation and measurement stages, see Figs. \ref{fig:MUB} to \ref{fig:random-bases}. We provided an easy to use code developed in Python to implement our algorithm, see \cite{pio-implementation}.

As interesting future lines of research, we pose the following list of open issues: (\emph{i}) Find an upper bound for fidelity reconstruction of Algorithm \ref{alg:pio1} as a function of errors and number of iterations; (\emph{ii}) Characterize the full set of quantum measurements for which Algorithm \ref{alg:pio1} converges in a single iteration; (\emph{iii}) Extend our method to quantum process tomography.\bigskip

\textbf{Acknowledgements}

It is a pleasure to thank Gustavo Ca\~nas C\'ardona, Zdenek Hradil, Felix Huber, Santiago G\'omez L\'opez, Kamil Korzekwa, Andrew Scott, Oliver Reardon-Smith, Stephen Walborn, Andreas Winter and Karol \.{Z}yczkowski for valuable comments.  DG and DU are supported by Grant FONDECYT Iniciaci\'{o}n number 11180474, Chile. DU also acknowledges support from Project ANT1956, Universidad de Antofagasta,
Chile. GS acknowledges support from the Government of Spain (FIS2020-TRANQI and Severo Ochoa CEX2019-000910-S), Fundació Cellex, Fundació Mir-Puig, Generalitat de Catalunya (CERCA, AGAUR SGR 1381) and the EU project QRANGE. This work was supported by MINEDUC-UA project, code ANT 1856.

As regards to the authorship of the different sections, DUC and DG provided both the theoretical background as well as the new mathematical results, whereas DUC and GS contributed with numerical simulations.
\appendix

\section{Proof of results}\label{proofs}
In this section we provide the proofs of all our results.

\subsection{Algorithm for quantum state estimation}

\textbf{Proposition II.1}
\emph{The following properties hold for any POVM $\{E_i\}_{i\leq m}$ and any $\rho$ acting on $\mathcal{H}_d$:
\begin{enumerate}
\item Imposition of physical information: $\mathrm{Tr}[T^{p_i}_{E_i}(\rho)E_i]=p_i.$
\item Composition: $T^{p_j}_{E_j}\circ T^{p_i}_{E_i}(\rho)=T^{p_i}_{E_i}(\rho)+T^{p_j}_{E_j}(\rho)-\rho-\bigl(p_i-\mathrm{Tr}(\rho E_i)\bigr)\mathrm{Tr}(E_iE_j)E_j/\mathrm{Tr}(E_j)^2.$
\item Non-expansiveness: $\mathfrak{D}(T^{p_j}_{E_j}(\rho),T^{p_j}_{E_j}(\sigma))\leq\mathfrak{D}(\rho,\sigma).$
\end{enumerate}
}
\begin{proof}
Items \emph{1} and \emph{2} easily arise from Definition \ref{def:pio}. In order to show the non-expansiveness stated in Item \emph{3}, let us apply Definition \ref{def:pio} to two states $\rho$ and $\sigma$, belonging to $\mathcal{H}_d$, i.e.
\begin{equation}\label{pio1Ap}
T^{p_i}_{E_i}(\rho)=\rho+\frac{(p_i-\mathrm{Tr}[\rho E_i])E_i}{\mathrm{Tr}(E_i^2)},
\end{equation}
\begin{equation}\label{pio2Ap}
T^{p_i}_{E_i}(\sigma)=\sigma +\frac{(p_i-\mathrm{Tr}[\sigma E_i])E_i}{\mathrm{Tr}(E_i^2)}.
\end{equation}
Subtracting \eqref{pio1Ap} from \eqref{pio2Ap}
\begin{equation}
T_{E_i}(\rho) - T_{E_i}(\sigma) = (\rho - \sigma) -  \dfrac{\mathrm{Tr}[(\rho - \sigma) E_i]E_i}{\mathrm{Tr}(E_i^2)},
\end{equation}
where we dropped the upper index $p_i$ from $T^{p_i}_{E_i}$. Now, let us compute $$\mathfrak{D}(T_{E_j}(\rho),T_{E_j}(\sigma))^2 = \mathrm{Tr}\bigl[ \bigl( T_{E_i}(\rho) - T_{E_i}(\sigma) \bigr)\bigl( T_{E_i}(\rho) - T_{E_i}(\sigma) \bigr)^{\dagger} \bigr].$$ Thus,
\begin{eqnarray}
\mathfrak{D}(T_{E_j}(\rho),T_{E_j}(\sigma))^2 &=& \mathfrak{D}(\rho, \sigma)^2 -  2\dfrac{\mathrm{Tr}[(\rho - \sigma) E_i]\mathrm{Tr}[(\rho - \sigma) E_i]}{\mathrm{Tr}(E_i^2)} + \dfrac{\bigl( \mathrm{Tr}[(\rho - \sigma) E_i]\bigr)^2 \mathrm{Tr}(E_i^2)}{\bigl( \mathrm{Tr}(E_i^2) \bigr)^2} \nonumber  \\
 &=&\mathfrak{D}(\rho, \sigma)^2 -  \dfrac{\bigl( \mathrm{Tr}[(\rho - \sigma) E_i]\bigr)^2}{\mathrm{Tr}(E_i^2)},
\end{eqnarray}
where $\mathfrak{D}(\rho, \sigma)^2 = \mathrm{Tr}\bigl[(\rho - \sigma)(\rho - \sigma)^{\dagger} \bigr]$. Therefore, $\mathfrak{D}(T_{E_j}(\rho),T_{E_j}(\sigma)) \leq \mathfrak{D}(\rho, \sigma)$ and item \emph{3} holds. 
\end{proof}

\textbf{Theorem III.1}
\emph{Let $A_1,\dots,A_{\ell}$ be a set of  informationally complete POVMs acting on a Hilbert space $\mathcal{H}_d$, associated to a compatible set of probability distributions $\vec{p_1},\dots,\vec{p_{\ell}}$. Therefore, Algorithm \ref{alg:pio1} converges to the unique solution to the quantum state tomography problem.}
\begin{proof}
First, from item \emph{1} in Prop. \ref{prop:pioproperties} the generator state $\rho_{gen}$ is a fixed point of each imposition operator $\mathcal{T}_{A_i}$, for every chosen POVM measurement $A_1,\dots,A_{\ell}$. Hence, $\rho_{gen}$ is a fixed point of the composition of all involved operators. Moreover, this fixed point is unique, as there is no other quantum state having the same probability distributions for the considered measurements, as $A_1,\dots,A_{\ell}$ are informationally complete. Here, we are assuming error-free probability distributions. Finally, convergence of our sequences is guaranteed by the alternating projections method developed by Halperin, which states that successive iterations of non-expansive projections converge to a common fixed point of the involved maps, see Theorem 1 in \cite{H67}.
\end{proof}



\subsection{Single-step convergence}

\textbf{Proposition III.1}
\emph{Let $\mathcal{T}_A$ and $\mathcal{T}_B$ be physical imposition operators associated to two mutually unbiased bases $A$ and $B$, for $n$ qudit systems. Therefore
\begin{equation}
\mathcal{T}_A\circ \mathcal{T}_B=\mathcal{T}_A+\mathcal{T}_B-\mathbb{I}.
\end{equation}
In particular, notice that $\mathcal{T}_A$ and $\mathcal{T}_B$ commute.}

\begin{proof}
First, it is simple to show that  $\mathcal{T}_A(\rho)=\rho_0+\sum_{j=0}^{m_A-1}\Pi_j(\rho-\rho_0)\Pi_j$ for any PVM $A$, where $\Pi_j=E_j$ are the subnormalized rank-one PVM elements. Thus, we have
\begin{eqnarray*}
T_B\circ T_A(\rho_0)&=&\rho_0+\sum_{j=0}^{m_A-1}\Pi^A_j(\rho-\rho_0)\Pi^A_j+\sum_{k=0}^{m_B-1}\Pi^B_k\left[\rho-\left(\rho_0+\sum_{j=0}^{m_A-1}\Pi^A_j(\rho-\rho_0)\Pi^A_j\right)\right]\Pi^B_k\\
&=&\rho_0+\sum_{j=0}^{m_A-1}\Pi^A_j(\rho-\rho_0)\Pi^A_j+\sum_{k=0}^{m_B-1}\Pi^B_k(\rho-\rho_0)\Pi^B_k+\sum_{j,k}\Pi^B_k\Pi^A_j(\rho-\rho_0)\Pi^A_j\Pi^B_k
\end{eqnarray*}
On the other hand,
\begin{eqnarray*}
\sum_{j,k}\Pi^B_k\Pi^A_j(\rho-\rho_0)\Pi^A_j\Pi^B_k&=&\sum_{j,k}\mathrm{Tr}(\Pi^A_j\Pi^B_k)\mathrm{Tr}\bigl((\rho-\rho_0)\Pi^A_j\bigr)\Pi^B_k\\
&=&\gamma(A,B)\sum_{j,k}\mathrm{Tr}\bigl((\rho-\rho_0)\Pi^A_j\bigr)\Pi^B_k\\
&=&\gamma(A,B)\mathrm{Tr}(\rho-\rho_0)\\
&=&0.
\end{eqnarray*}
Therefore, we have
\begin{eqnarray}
T_B\circ T_A(\rho_0)&=&\rho_0+\sum_{j=0}^{m_A-1}\Pi^A_j(\rho-\rho_0)\Pi^A_j+\sum_{k=0}^{m_B-1}\Pi^B_k(\rho-\rho_0)\Pi^B_k\\
&=&T_A(\rho_0)+T_B(\rho_0)-\rho_0,
\end{eqnarray}
for any initial state $\rho_0$. So, we have 
$T_B\circ T_A=T_A\circ T_B=T_A+T_B-\mathbb{I}$.
\end{proof}

\textbf{Proposition III.2}
\emph{Algorithm \ref{alg:pio1} converges in a single iteration to the unique solution of the quantum state tomography problem for product of generalized Pauli operators and also for $d+1$ mutually unbiased bases, in any prime power dimension $d$.}
\begin{proof}
For generalized Pauli operators, commutativity of imposition operators comes from orthogonality condition $\mathrm{Tr}(E_iE_j)$, see item \emph{2} in Prop. \ref{prop:pioproperties}. Thus, we have
\begin{eqnarray}\label{seq}
\rho_n&=&(T_{E_{d^2}}\circ\cdots\circ T_{E_1})^n(\rho_0)\nonumber\\
&=&T^n_{E_{d^2}}\circ\cdots\circ T^n_{E_1}(\rho_0)\nonumber\\
&=&T_{E_{d^2}}\circ\cdots\circ T_{E_1}(\rho_0),
\end{eqnarray}
where the second step considers commutativity and the last step the fact that every $T_j$, $j=1,\dots,d+1$ is a projection. On the other hand, from Theorem \ref{thm:convergence} we know that $\rho_n\rightarrow\rho_{gen}$ when $n\rightarrow\infty$, for any generator state $\rho_{gen}$. From combining this result with (\ref{seq}) we have 
\begin{equation}
T_{E_{d^2}}\circ\cdots\circ T_{E_1}(\rho_0)=\rho_{gen},
\end{equation}
for any seed $\rho_0$ and any generator state $\rho_{gen}$, in any prime power dimension $d$.

For MUB the result holds in the same way, where commutativity between the associated imposition operators associated to every PVM  arises from see Prop. \ref{MUBcommute}. 
\end{proof}

\section{An additional model of errors for the measurement process}\label{sec:simulations-Gaussian_Noisse}
\begin{figure}
      \includegraphics[scale=0.6]{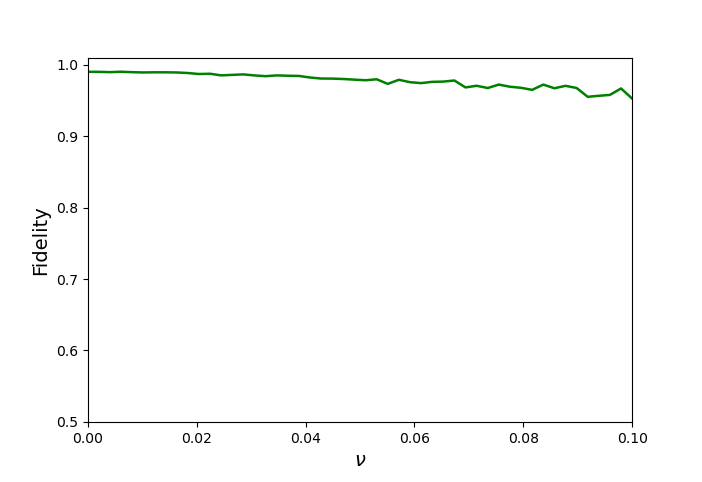}
     \caption{A new error model for the measurement process, which considers a Gaussian perturbation of the spin direction to be measured together with finite statistics errors. Fidelity is averaged over 100 trials, having a randomly chosen generator state $\rho_{gen}$ each. Measurement statistics are estimated from 200 identical copies of the target state, where we consider eigenbases of spin $1/2$ observables in three orthogonal directions.} 
     \label{Fig6}
\end{figure}
Along the work, we implemented simulations considering errors in both state preparation and those arising from finite statistics. In this section, we consider an additional source of errors in the measurement process. Specifically, we consider errors in the measurement apparatus, which is modeled by adding Gaussian perturbations in the direction of spin observables. In figure \ref{Fig6}, we show fidelity for quantum state reconstruction for a spin $1/2$ particle from three spin observables along orthogonal directions. For the Gaussian noise model, such directions are affected by a Gaussian probability distribution having standard deviation $\nu$, centered in the ideally expected direction. That is, we consider the Gaussian probability distribution $p (x) \propto e^{-(x - \mu)^2/2\nu^2 }$ with $\mu = 0$, for entries of a spin direction $n$, associated to the observable $S = \vec{n}\cdot\vec{\sigma}$, where $\vec{\sigma}=(\sigma_x,\sigma_y,\sigma_z)$ is a vector composed by the three Pauli matrices. The amplitude of fluctuations can be controlled by adjusting the standard deviation $\nu$.

\end{document}